# Static critical behavior of the ferromagnetic transition in $LaMnO_{3.14}$ manganite

R.S. Freitas[a], C. Haetinger[b], P. Pureur[b], J.A. Alonso[c], and L. Ghivelder[a]

[a] Instituto de Física, Universidade Federal do Rio de Janeiro, c.p. 68528, Rio de Janeiro, 21945-970, RJ, Brazil

[b] Instituto de Física, Universidade Federal do Rio Grande do Sul, c.p. 15051, Porto Alegre, 91501-970, RS, Brazil

[c] Instituto de Ciencia de Materiales de Madrid, C.S.I.C., Cantoblanco, 28049 Madrid, Spain

The ferromagnetic phase transition in $LaMnO_{3.14}$ is investigated by measuring the dc magnetization as a function of magnetic field and temperature. Modified Arrott plot and Kouvel Fisher analysis yield estimates for the critical exponents $\beta$, and $\gamma$, with values between that predicted for the Heisenberg model and mean field theory. At low fields we found an anomalous small value of $\beta$, indicating that the critical behavior is influenced by the range of magnetic fields used.

The discovery of the colossal magnetoresistance (CMR) in hole-doped $LaMnO_3$ perovskites has attracted renewed interest in this class of materials, and numerous papers appeared in which the paramagnetic-ferromagnetic phase transition was investigated [1-5]. In the study of the critical behavior associated with this transition there is considerable disagreement among the experimental estimates for the exponent $\beta$, associated with the temperature dependence of spontaneous magnetization. Lofland *et al.* [1] found for $La_{0.7}Sr_{0.3}MnO_3$ a value of $\beta = 0.45$, indicating a mean-field-like behavior. Ghosh *et al.* [2], studying the same compound, found a rather low value, $\beta = 0.37$, which is close to that predicted by the Heisenberg model. For the $La_{0.7}Ca_{0.3}MnO_3$ system the situation is more controversial. While Heffner *et al.* [3] performed muon spin relaxation measurements which points to a second order transition, neutron diffraction [4] and magnetization studies [5] indicate that the transition is discontinuous.

In the present report we have investigated the critical behavior at the paramagnetic-ferromagnetic transition of a $LaMnO_{3+\delta}$ ($\delta = 0.14$) polycrystalline sample. Stoichiometric $LaMnO_3$ is antiferromagnetic at low temperatures. The oxygen excess $\delta$ results in La and Mn vacancies that induces a $Mn^{3+}$-$Mn^{4+}$ mixed-valence state, responsible for ferromagnetism via double-exchange mechanism. The sample was prepared in polycrystalline form by a citrate technique, as described elsewhere [6]. The determination of $\delta = 0.134$, corresponding to 27% of $Mn^{4+}$, was performed by thermogravimetric analysis. The final material was characterized by x-ray diffraction. Detailed dc magnetization measurements were performed in a commercial magnetometer. The results are corrected for the demagnetization factor of the sample.

Figure 1 shows magnetization isotherms in the form of the modified Arrot plots, $M^{1/\beta}$ vs $(H/M)^{1/\gamma}$, based on the Arrot-Noakes equation of state [7]. This plot allowed an estimation of the critical exponents when the isotherms measured at different temperatures close the $T_c$ are linear and parallel to one another. This condition is satisfied in the high-field region (0.5 to 4 kOe) with $\beta = 0.415$ and $\gamma = 1.47$. The straight-line at $T = T_c$ passes through the origin, which gives $T_c = 141$ K. As a further test of these values, the results are compared with the prediction of a scaling hypothesis, in which $M/|\varepsilon|^{\beta}$ plotted as a function of $H/|\varepsilon|^{(\beta+\gamma)}$, where $\varepsilon = (T-T_c)/T_c$, should give two different curves, one for $T > T_c$ and another for $T<T_c$. This is indeed observed in the scaled data plotted in Fig. 2, indicating that the obtained values of $T_c$, $\beta$, and $\gamma$ are reliable. The value of $\beta = 0.415$ is between that predicted for the Heisenberg model ($\beta = 0.365$) and the one predicted by the mean field theory ($\beta = 0.5$).

As an additional method for obtaining the critical exponents we have used the Kouvel-Fisher method [8]; defining $M_s$ as the spontaneous magnetization and $\chi_0$ as the initial susceptibility, this method suggests that the quantity $M_s[dM_s/dT]^{-1}$ and $\chi_0[d\chi_0/dT]^{-1}$ plotted against temperature yield straight-lines, with slopes $(1/\beta)$ and $(-1/\gamma)$, respectively. The intercepts of these lines on the temperature axes are equal to $T_c$. The $M_s(T)$ and $\chi_0$ (T) values are obtained from the linear extrapolation to $(H/M)^{1/\gamma}=0$ and $M^{1/\beta}=0$ of the high-field straight line portions of the isotherms on the modified Arrot plot (Fig.1). With these values, the critical exponents obtained are $\beta = 0.42$, $\gamma = 1.34$, and $T_c = 141$ K. These values are similar those found directly from the modified Arrott plot.

In order to investigate if these results are influenced by the magnetic field we have used the Kouvel-Fisher method with values of magnetization and susceptibility measured at low fields, H = 10, 30, 50, 70, 100, 150 Oe. Figure 3 shows the data for H = 70 Oe. The linear fits for all fields gives virtually the same results, with an average of $\beta = 0.095$, $\gamma = 1.47$ and $T_c = 141$ K. The very small value of $\beta$ indicates that at low fields the ferromagnetic transition can be interpreted as being nearly first order. This interpretation is based on the fact that $\beta$ values close to zero indicate a tendency towards a discontinuous transition. The $\gamma$ value is identical to that obtained from the modified Arrot plot. We explain this fact by noting that this exponent is obtained from magnetization data just above Tc, and therefore it is less affected by the applied field. The influence of the demagnetization factor in the low field results is not relevant, since the demagnetizing field is proportional to the magnetization, which is small near Tc. This was further investigated by measuring the same sample after cutting it in various sizes, and similar values for the critical exponents were obtained.

In summary, we have studied the critical behavior of $LaMnO_{3.14}$ and determined the values of $T_c$, $\beta$, $\gamma$. Furthermore, the very small value of $\beta$ at low fields indicate that the ferromagnetic transition may have a weakly first-order character.

References

[1] S.E. Lofland, V. Ray, P.H. Kim, S.M. Bhagat, M.A. Manheimer, and S.D. Tyagi, Phys. Rev. B 55 (1997) 2749.
[2] K. Ghosh, C.J. Lobb, R.L. Greene, S.G. Karabashev, D.A. Shulyatev, A.A. Arsenov, and Y.M. Mukovskiy, Phys. Rev. Lett. 81 (1998) 4740.
[3] R.H. Heffner, L.P. Le, M.F. Hundley, J.J. Neumeier, G.M. Luke, K. Kojima, B. Nachumi, Y.J. Uemura, D.E. MacLaughlin, and S-W. Cheong, Phys. Rev. Lett. 77 (1996) 1869.

[4] J.W. Lynn, R.W. Erwin, J.A. Borchers, Q. Huang, A. Santoro, J-L. Peng, and Z.Y. Li, Phys. Rev. Lett. 76 (1996) 4046.
[5] J. Mira, J. Rivas, F. Rivadulla, C. Vázquez-Vázquez, and M.A. López-Quintela, Phys. Rev. B 60 (1999) 2929.
[6] J.A. Alonso, M.J. Martinez-Lope, M.T. Casais, J.L Macmanus-Driscoll, P.S.I.P.N. de Silva, and L.F. Cohen, J. Mater. Chem. 7 (1997) 2139.
[7] A. Arrot and J.E. Noakes, Phys. Rev. Lett. 19 (1967) 786.
[8] J.S. Kouvel, M.E. Fisher, Phys. Rev. A 136 (1964) 1626.

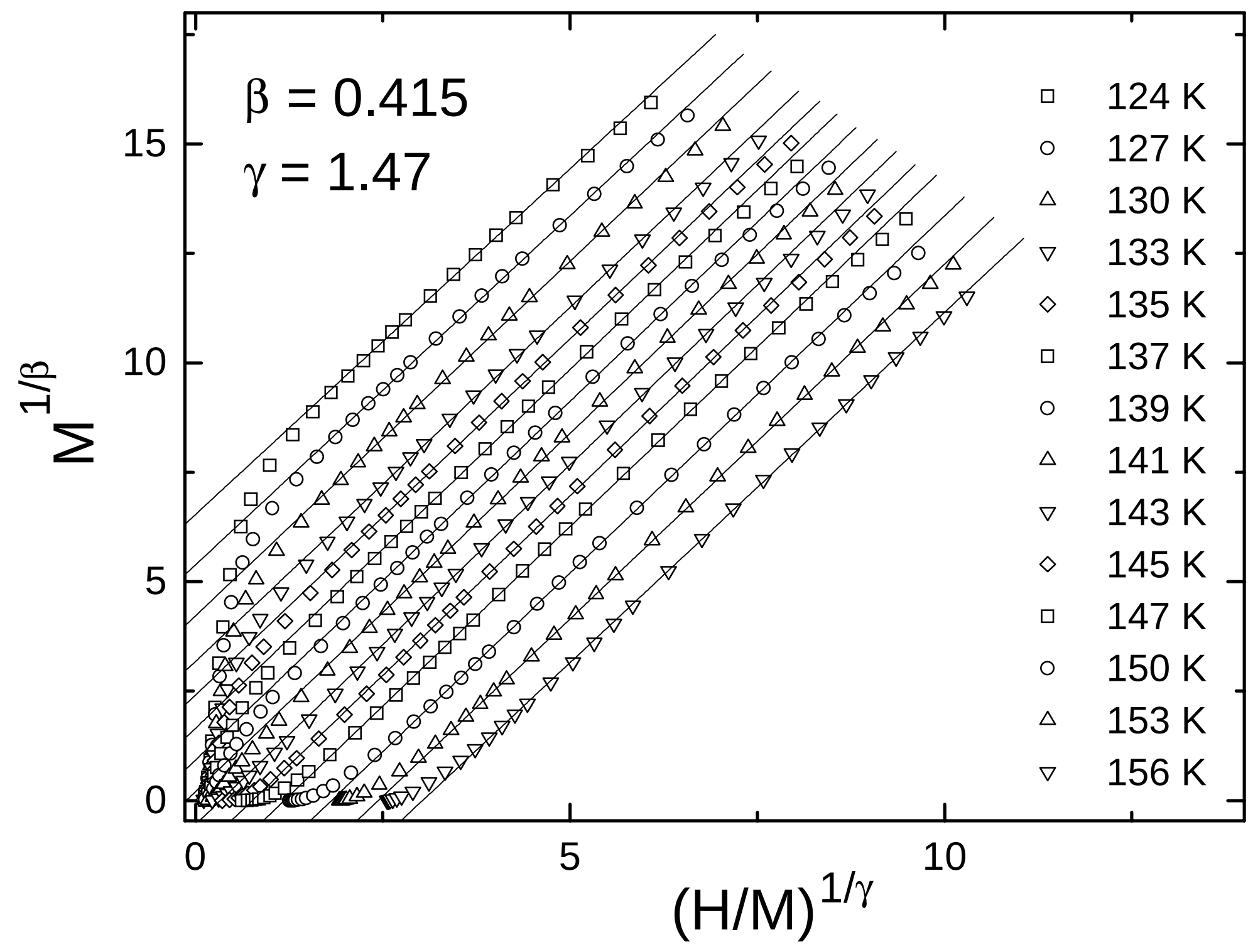

Fig. 1 - Modified Arrott plot isotherms on $LaMnO_{3.14}$.

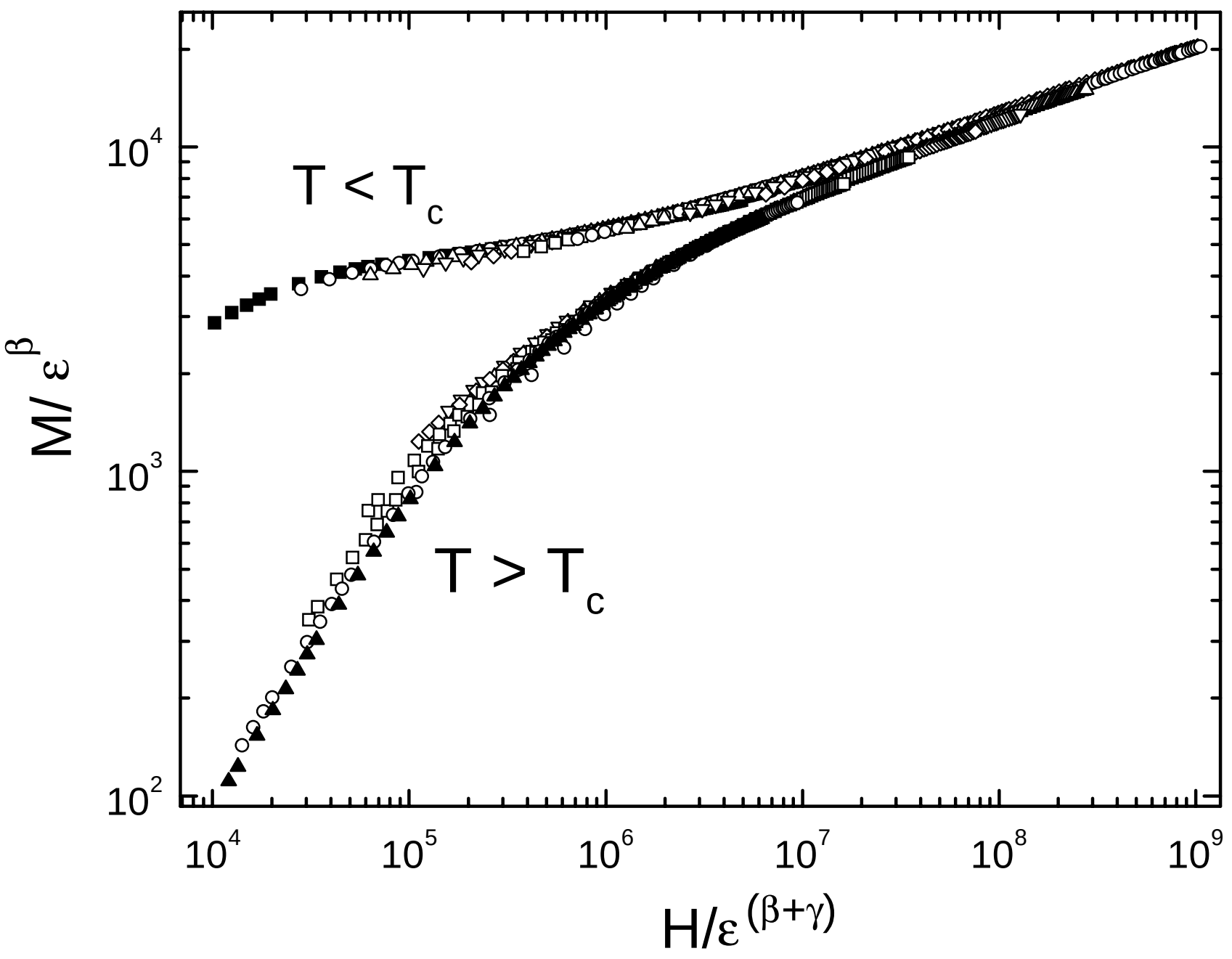

Fig. 2 - Scaling plot on a log scale, as explained in the text.

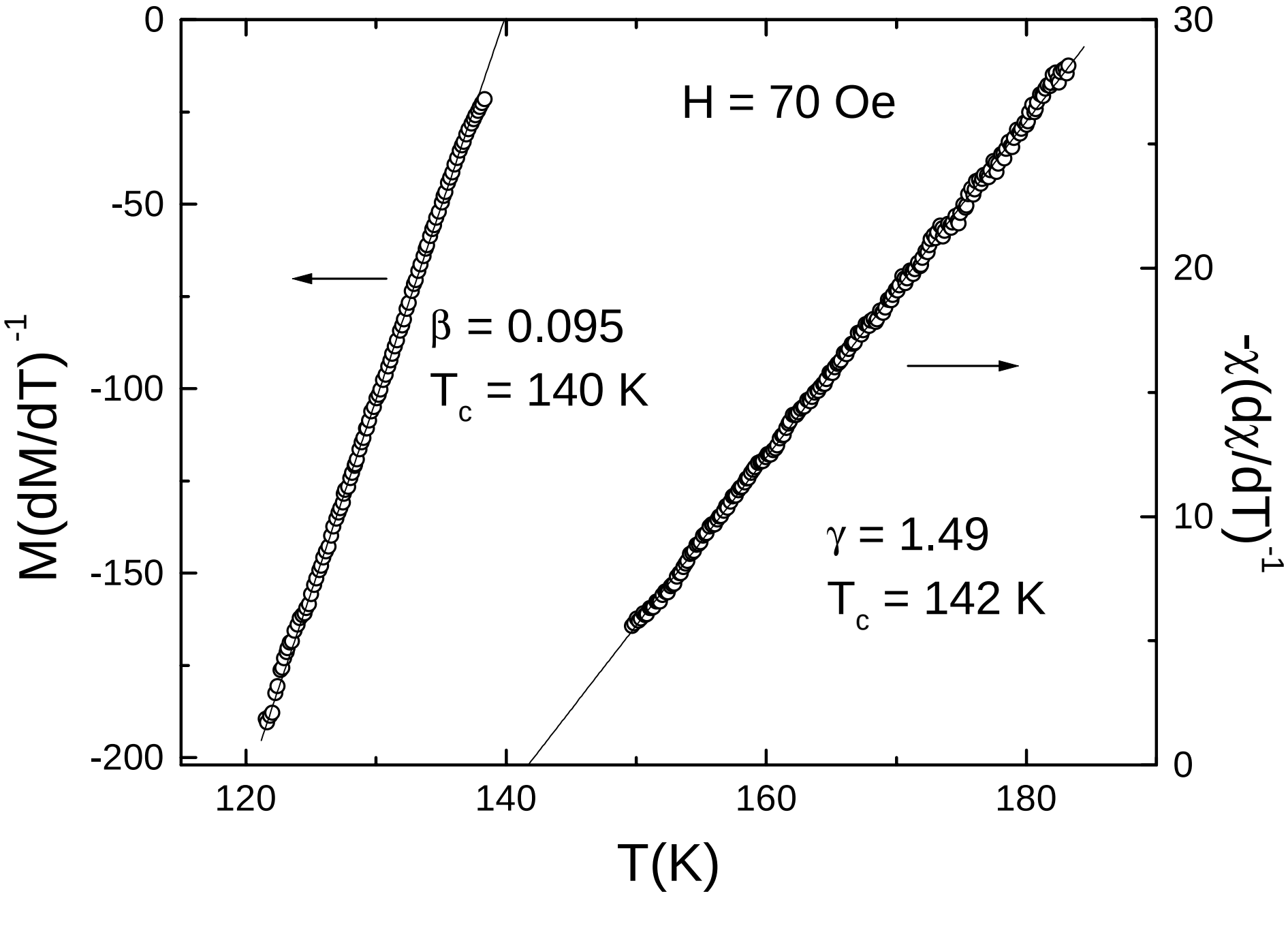

Fig. 3 - Kouvel-Fisher plot for the magnetization and susceptibility data at H = 70 Oe.